\documentclass{elsart3}
\usepackage{graphicx}
\newcommand{\bnmr}{$\beta$-NMR}
\newcommand{\bnqr}{$\beta$-NQR}
\newcommand{\STO}{SrTiO$_3$}
\newcommand{\Sap}{Al$_2$O$_3$}
\newcommand{\SRO}{Sr$_2$RuO$_4$}

\begin{document}

\begin{frontmatter}
\title{Beta-Detected NQR in Zero Field with a Low Energy Beam of \nuc{8}{Li}$^+$}
\author[TR]{Z. Salman\corauthref{cor}}
\corauth[cor]{Tel. +1 604 222 1047 x 6588,
Fax: +1 604 222 1074,
email: zaher@triumf.ca} 
\author[UBCP]{R.F. Kiefl}
\author[AB]{K.H. Chow}
\author[UBCC]{W.A. MacFarlane}
\author[TR]{S.R. Kreitzman}
\author[TR]{D.J. Arseneau}
\author[TR]{S. Daviel}
\author[TR]{C.D.P. Levy}
\author[KY]{Y. Maeno}
\author[TR]{R. Poutissou} 
\address[TR]{TRIUMF, 4004 Wesbrook Mall, Vancouver, BC, V6T 2A3,
  Canada}
\address[UBCP]{Department of Physics and Astronomy, University of British Columbia, Vancouver, BC, V6T 1Z1, Canada}
\address[AB]{Department of Physics, University of Alberta, Edmonton,
  AB, T6G 2J1, Canada}
\address[UBCC]{Department of Chemistry, University of British
  Columbia, Vancouver, BC, V6T 1Z1, Canada}
\address[KY]{Department of Physics, Kyoto University, Kyoto 606-8502, Japan}

\begin{abstract}
Beta-detected nuclear quadrupole resonances (\bnqr) at
zero field are observed using a beam of low energy highly polarized
radioactive \nuc{8}{Li}$^+$. The resonances were detected in \STO,
\Sap\ and \SRO\ single crystals by monitoring the beta-decay
anisotropy as a function of a small audio frequency magnetic
field. The resonances show clearly that \nuc{8}{Li} occupies one site
with non-cubic symmetry in \STO, two in \Sap\ and three sites in
\SRO. The resonance amplitude and width are surprisingly large
compared to the values expected from transitions between the $|\pm
2\rangle \leftrightarrow |\pm 1\rangle$ spin states, indicating a
significant mixing between the $|\pm m \rangle$ quadrupolar split
levels.
\end{abstract}

\end{frontmatter}

\section{Introduction}
Recently, we have constructed two spectrometers for beta-detected
nuclear magnetic resonance \cite{Morris04PRL} (\bnmr) and nuclear
quadrupole resonance \cite{Salman04PRB} (\bnqr), using a low energy
highly polarized \nuc{8}{Li}$^{+}$ beam. We report here on the first
results obtained from the \bnqr\ spectrometer at zero applied magnetic
field. The \bnqr\ spectra were obtained for \nuc{8}{Li} implanted into
single crystals of \STO, \Sap\ and \SRO. The ability to perform
measurements in zero applied field has many potential applications in
studies of magnetism and superconductivity. It is also remarkable that
the resonances are narrow (few kHz) and easily observed at acoustic
frequencies.

This work is intended as an initial characterization of the spectra of
\nuc{8}{Li} in \STO, \Sap, and \SRO. \STO\ is probably the
best-studied perovskite transition metal oxide. It is interesting for
its prototypical soft mode structural phase transition ($\sim 105$~K)
\cite{Cowley96PTSL}, its ferroelectric properties
\cite{Bednorz84PRL,Itoh99PRL}, and as a high dielectric constant layer
in heterostructures based on Si \cite{McKee98PRL}. Both \STO\ and
\Sap\ are important substrate materials for thin films, and therefore
it is important to understand the behavior of \nuc{8}{Li} in these
substrates for future studies. \SRO\ is an unconventional spin-triplet
superconductor ($T_c=1.5$~K) as demonstrated by NMR Knight shift
measurements \cite{Ishida98N}. In its normal state it also exhibits a
highly correlated metallic behavior which can be described as a
quasi-two-dimensional Fermi liquid \cite{Mackenzie96PRL,Maeno97JPSJ}.

The experiment was performed at the TRIUMF ISAC facility in the new
\bnqr\ spectrometer. In this experiment a highly polarized beam of
\nuc{8}{Li} is implanted in the sample with energy of $30$~keV. At
this energy the average implantation depth is $\sim 200$~nm. A
linearly polarized oscillating magnetic field $B_1$ is applied
perpendicular to the initial nuclear spin polarization. A resonant
loss of the nuclear beta decay asymmetry of \nuc{8}{Li} occurs when
the frequency of the oscillating field $\nu$ matches the nuclear
energy spin level splitting. A more detailed description of the \bnqr\
spectrometer and of the beta detected nuclear resonance technique used
can be found in Ref.\cite{Salman04PRB} and references therein. Once
the development of this spectrometer is completed it will have the
capability to reduce the beam energy to $100$~eV, allowing for depth
profiling measurements on a nm scale in zero applied magnetic field.

\begin{figure}[h]
\centering
\includegraphics[width=5.0cm]{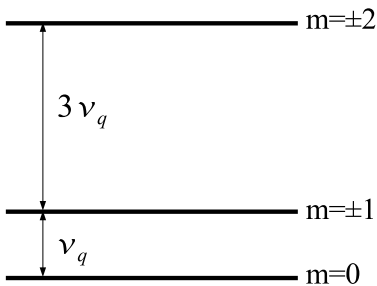}
\caption{The energy levels diagram for the Hamiltonian (\ref{Hq}).} \label{ELevel}
\end{figure}

The Hamiltonian for the implanted \nuc{8}{Li} at zero field in the
presence of an axially symmetric electric field gradient (EFG) is
\cite{Slichter}
\begin{equation} \label{Hq}
H_q= h\nu_q[{I_z^{2}-2}] 
\end{equation}
where $\nu_q=e^{2}qQ/8$, $eq=V_{zz}$ is the electric field gradient at
the \nuc{8}{Li} site, and $Q$ is the electric quadrupole moment of the
nucleus. Therefore, when $q \ne 0$ even at zero magnetic field a
splitting between the nuclear spin sub-levels $|m \rangle $ is present
(see Fig.~\ref{ELevel}).
In this case two resonance frequencies are possible, one at
$\nu=\nu_q$ due to the transitions $|\pm 1\rangle \leftrightarrow |0
\rangle$, and another at $\nu=3\nu_q$ due to $|\pm 2 \rangle
\leftrightarrow |\pm 1\rangle$ transitions.

\section{Results}
\begin{figure}[h]
\centering
\includegraphics[width=\columnwidth]{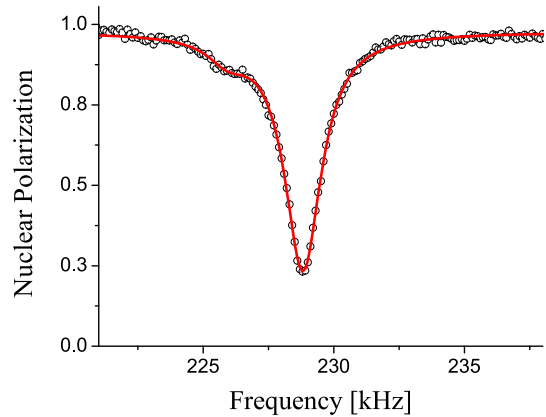}
\caption{The \bnqr\ spectrum in \STO\ at room temperature and zero
  applied field} \label{STO}
\end{figure}

\bnqr\ spectra were collected on an epitaxially polished $\langle 100
\rangle$ single crystal of \STO\ (Applied Crystal Technologies). We
have established earlier that the \nuc{8}{Li} occupies the face
centered cite in \STO\ \cite{Salman04PRB,MacFarlane03PB3}, and
therefore experiences a non-vanishing EFG. Indeed a large and sharp
NQR resonance was observed at $3\nu_q=228.81(2)$ kHz with width $1.7$
kHz corresponding to the $|\pm 2\rangle \leftrightarrow |\pm 1\rangle$
transitions (see Fig. \ref{STO}), whereas the resonance near $\nu_q$
was barely visible, indicating very low probability of \nuc{8}{Li} in
the $|\pm 1\rangle$ spin states as expected from the high polarization
of the \nuc{8}{Li} beam.
Note the resonance in Fig. \ref{STO} is slightly asymmetric.  The
solid curve is a fit assuming two overlapping lines.  The smaller
amplitude line occurs at a slightly lower frequency $\nu=225.9(2)$ kHz
and has a width of about $1.9(1)$ kHz which is similar to the higher
frequency line at $\nu=228.8$ kHz. 

\begin{figure}[h]
\centering
\includegraphics[width=\columnwidth]{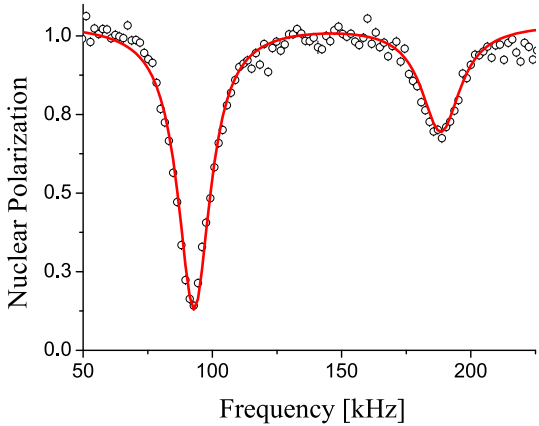}
\caption{The \bnqr\ spectrum in \Sap\ at room temperature and zero
  applied field.} \label{Sapphire}
\end{figure}

In \Sap\ (epitaxially polished, Honeywell) and \SRO\ (freshly cleaved)
the beta decay asymmetry was found to be zero when the \nuc{8}{Li} was
implanted with its nuclear polarization perpendicular to the
c-axis. However, when the sample was rotated by $45^o$ significant
asymmetry was observed, a clear indication that the EFG experienced
by the implanted \nuc{8}{Li} is along the c-axis in these crystals. In
\Sap\ the spectrum (see Fig.~\ref{Sapphire}) shows two distinct
resonance lines at $\nu=92.94(8)$ and $188.9(3)$~kHz, with widths of
$15.1(3)$ and $22.5(9)$~kHz respectively. The two resonances are due
to at least two inequivalent \nuc{8}{Li} sites.
However, note that the widths of the lines in \Sap\ are significantly
larger than that observed in \STO, likely due to a larger distribution
of $\nu_q$, which may be caused by multiple \nuc{8}{Li} sites with
very similar values of $\nu_q$. This may be expected considering the
complexity of the lattice structure of \Sap.

\begin{figure}[h]
\centering
\includegraphics[width=\columnwidth]{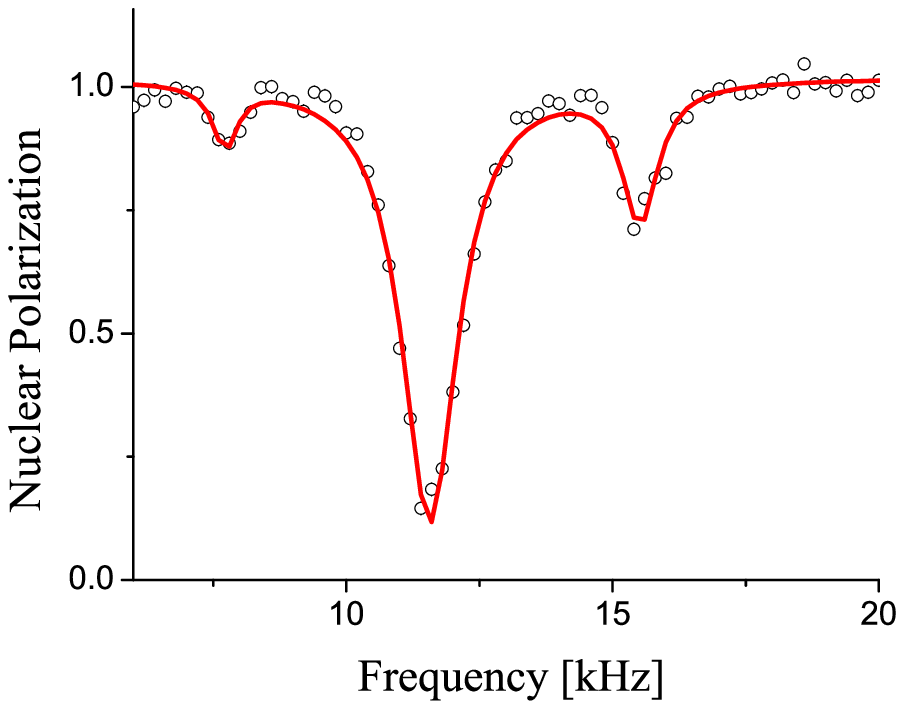}
\caption{The \bnqr\ spectrum in \SRO\ at room temperature and zero
  applied field} \label{SRO}
\end{figure}

The \bnqr\ spectrum in \SRO\ is shown in Fig.~\ref{SRO} at room
temperature and zero field, where three sharp and separated resonances
are observed at $\nu=7.72(4)$, $11.57(1)$ and $15.51(2)$~kHz, with
corresponding width of $0.55(13)$, $1.28(3)$ and $0.83(7)$ kHz,
indicating three well defined \nuc{8}{Li} lattice sites in this
tetragonal material. The resonances here are even sharper than in
\STO, reflecting the high quality of this crystal.

In all three cases presented, the amplitude of the resonances are much
larger than expected. The optical pumping method used to generate the
nuclear polarization of the \nuc{8}{Li} beam produces polarization as
high as $\sim 70 \%$ along the measurement axis ($z$). This
polarization can be defined as
\begin{equation} \label{polarization}
P_z=\frac{1}{2} \sum_{m=-2}^{m=+2} p_m m
\end{equation}
where $p_m$ is the probability that the $|m\rangle$ state is
occupied. Realistic values for these probabilities are 
\[ \{p_{+2},p_{+1},\cdots,p_{-2}\}=\{0.65,0.2,0.15,0,0\}. \]
When the frequency of $B_1$ is $\nu=3\nu_q$, and assuming sufficient
 power to saturate the transition, the probability that the
 $|+2\rangle$ or $|+1\rangle$ are occupied becomes equal. This implies
 that the polarization is reduced from its initial value, $0.75$, to
 $0.6375$ with probabilities $\{0.425,0.425,0.15,0,0\}$, i.e.  $15 \%$
 loss in the polarization or asymmetry. Our measurements show a
 considerably larger amplitude. This enhancement can be explained by
 an additional term in the Hamiltonian, which mixes the different
 sub-levels $|m\rangle$, and produces a larger effect by allowing
 transitions other than $|\pm 2 \rangle \leftrightarrow |\pm 1
 \rangle$. Such a term can be produced by a small stray magnetic field
 perpendicular to the EFG axis, or by non-axial terms in the EFG. In
 the case of \STO\ we concluded that the non-axial terms in the EFG
 are responsible for this enhancement \cite{Salman04PRB}. However,
 more measurements are required on \Sap\ and \SRO\ to identify the
 origin of the additional interaction responsible for the amplitude
 enhancement.

Interestingly, in addition to the large amplitudes, the sum of the
amplitudes in the case of \Sap\ and \SRO\ is larger than $1$, as seen
in Fig.~\ref{Sapphire} and \ref{SRO}. This is only possible if
\nuc{8}{Li} is diffusing between inequivalent sites. When the
diffusion rate is higher than the \nuc{8}{Li} decay rate, it would be
possible for a \nuc{8}{Li} particle which is initially {\em off
resonance} to move to an {\em on resonance} site, thus increasing the
measured amplitude. This will be verified by cooling the sample and
measuring the \bnqr\ spectra as a function of temperature, where one
expects the diffusion rate to slow down, and consequently the sum of
amplitudes to drop below $1$.

\section{Summary and Conclusion}
We have demonstrated that it is possible to carry out $\beta$ detected
nuclear quadrupole resonance using a beam of low energy highly
polarized \nuc{8}{Li}$^+$. Clear $\beta$-NQRs were observed in \STO,
\Sap, and \SRO\ indicating that the implanted Li adopts well defined
crystalline lattice sites. In contrast to $\mu^+$, the quadrupole
resonances provide a means of identifying the \nuc{8}{Li} site. 

There is evidence for small terms in spin Hamiltonian which lead to
mixing of the $| \pm m \rangle$ states and a dramatic enhancement of
the amplitude of the resonances at $3\nu_q$. The ability to perform
$\mu$SR in zero field has been used extensively in studies of magnetism
and and superconductors. We anticipate similar applications are
possible with \bnqr\ in studies of ultra-thin films and interfaces.
For example in superconductors it could be used to measure the
absolute value of the London penetration depth or to search for states
with broken time reversal symmetry. In semiconductors or ionic
compounds it can be used to study the diffusion and electronic
structure of isolated Li in reduced geometries.

\ack{This work was supported by the CIAR, NSERC and TRIUMF.  We thank Rahim
Abasalti and Bassam Hitti for technical support We also thank Laura
Greene for providing the SrTiO$_3$ sample.}

\newcommand{\noopsort}[1]{} \newcommand{\printfirst}[2]{#1}
  \newcommand{\singleletter}[1]{#1} \newcommand{\switchargs}[2]{#2#1}

\end{document}